\documentclass[12pt]{article}
\usepackage{graphicx}
\usepackage{amsmath}
\usepackage{amssymb}

\begin{document}

\title{\bf Width of the chaotic layer: maxima due to marginal resonances}

\author{Ivan~I.~Shevchenko\/\thanks{E-mail:~iis@gao.spb.ru} \\
Pulkovo Observatory of the Russian Academy of Sciences \\
Pulkovskoje ave.~65, St.Petersburg 196140, Russia}
\date{}

\maketitle


\begin{center}
Abstract
\end{center}

\noindent Modern theoretical methods for estimating the width of
the chaotic layer in presence of prominent marginal resonances are
considered in the perturbed pendulum model of nonlinear resonance.
The fields of applicability of these methods are explicitly and
precisely formulated. The comparative accuracy is investigated in
massive and long-run numerical experiments. It is shown that the
methods are naturally subdivided in classes applicable for
adiabatic and non-adiabatic cases of perturbation. It is
explicitly shown that the pendulum approximation of marginal
resonance works good in the non-adiabatic case. In this case, the
role of marginal resonances in determining the total layer width
is demonstrated to diminish with increasing the main parameter
$\lambda$ (equal to the ratio of the perturbation frequency to the
frequency of small-amplitude phase oscillations on the resonance).
Solely the ``bending effect'' is important in determining the
total amplitude of the energy deviations of the near-separatrix
motion at $\lambda \gtrsim 7$. In the adiabatic case, it is
demonstrated that the geometrical form of the separatrix cell can
be described analytically quite easily by means of using a
specific representation of the separatrix map. It is shown that
the non-adiabatic (and, to some extent, intermediary) case is most
actual, in comparison with the adiabatic one, for the physical or
technical applications that concern the energy jumps in the
near-separatrix chaotic motion.

\bigskip

\noindent Key words: Hamiltonian dynamics, chaotic dynamics,
chaotic layer, separatrix map.


\section{Introduction}

Width of the chaotic layers near separatrices of Hamiltonian
systems is subject to large jump-like perturbations (upon
variation of parameters) due to emergence of marginal resonances
at the borders of the layers \cite{C79,S98}. The width is maximal,
or close to maximal, when a prominent marginal resonance is
present at the border of the chaotic layer, and its separatrix
chaotic layer is in heteroclinic connection with the main layer.
The width is maximal when the two layers are on the brink of
heteroclinic disconnection.

Formulas for the maximal width and for critical values of
parameters were derived in \cite{S98} in the perturbed pendulum
model of nonlinear resonance. This is the model that naturally
arises in a wide range of problems of physics and celestial
mechanics, when interactions of nonlinear resonances are
considered \cite{C79}. The field of applicability of the theory
\cite{S98} is the realm of non-adiabatic chaos \cite[p.~813]{S08}.
In fact, the model \cite{S98} covers the realm of non-adiabatic
(and, to some extent, intermediary) chaos for any system described
by the separatrix map. As found in \cite{S08}, the border between
slow (adiabatic) and fast (non-adiabatic) chaos is rather sharp
and is located at $\lambda \approx 1/2$ for any system described
by the separatrix map, given by Eqs.~(2) in \cite{S08} (see
Eqs.~(\ref{sm1}) below). The quantity $\lambda$ is the
``adiabaticity parameter'' \cite{C79}, equal to the ratio of the
perturbation frequency to the frequency of small-amplitude phase
oscillations on the resonance.

The non-adiabatic approximation was adopted in \cite{S98} on
several reasons. The separatrix map theory, which forms the basis
for the subsequent theories of the near-separatrix energy jumps,
including \cite{S98}, \cite{S08}, and \cite{SM09}, was initially
developed and believed to be applicable in the non-adiabatic case
$\lambda \gg 1$, see \cite{C79,V02}. (Later on, in \cite{S00}, the
separatrix map theory was shown to be applicable in the adiabatic
domain. An analytical description of marginal resonances in this
domain was performed in \cite{V02} by the method of so-called
resonance invariants.) The study \cite{S98} was inherently aimed
at real applications (in celestial mechanics), where the adiabatic
limit was not actual. Besides, the boundary in $\lambda$ between
slow and fast chaos had not been yet identified, neither it had
been known to be sharp, until the boundary was revealed in
\cite{S08}. Of course, in a moderately adiabatic case, say, at
$\lambda$ as low as $0.1$, the theory \cite{S98} can be also used,
as a rough extrapolation.

\section{The model of nonlinear resonance and the separatrix map}
\label{mnr}

For the model of nonlinear resonance we take that adopted in
\cite{S98,S08}. Under general conditions~\cite{C77,C79,LL92}, a
model of nonlinear resonance is provided by the Hamiltonian of
nonlinear pendulum with periodic perturbations. A number of
problems on non-linear resonances in mechanics and physics is
described by the Hamiltonian

\vspace{-3mm}

\begin{equation}
H = {{{\cal G} p^2} \over 2} - {\cal F} \cos \varphi +
    a \cos(\varphi - \tau) + b \cos(\varphi + \tau)
\label{h}
\end{equation}

\noindent (see e.g. \cite{S00}). The first two terms in
Eq.~(\ref{h}) represent the Hamiltonian $H_0$ of the unperturbed
pendulum, and the last two terms represent the periodic
perturbations; $\varphi$ is the pendulum angle (the resonance
phase angle), $p$ is the momentum; $\tau$ is the phase angle of
perturbation: $\tau = \Omega t + \tau_0$, where $\Omega$ is the
perturbation frequency, and $\tau_0$ is the initial phase of the
perturbation. The quantities ${\cal F}$, ${\cal G}$, $a$, $b$ are
constants. We assume that ${\cal F} > 0$, ${\cal G} > 0$, and $a =
b$.

The so-called separatrix (or ``whisker'') map

\vspace{-3mm}

\begin{eqnarray}
& & w_{i+1} = w_i - W \sin \tau_i,  \nonumber \\
& & \tau_{i+1} = \tau_i +
                 \lambda \ln {32 \over \vert w_{i+1} \vert}
                 \ \ \ (\mbox{mod } 2 \pi),
\label{sm}
\end{eqnarray}

\noindent written in the present form and explored in
\cite{C77,C78,C79} and first introduced (in implicit form) in
\cite{ZF65}, describes the motion in the vicinity of the
separatrices of Hamiltonian~(\ref{h}). The quantity $w$ denotes
the relative (with respect to the unperturbed separatrix value)
pendulum energy $w \equiv {H_0 \over {\cal F}} - 1$; $\tau$ is the
phase angle of perturbation (the same as in Eq.~(\ref{h})).

The constants $\lambda$ and $W$ are the two basic parameters. The
parameter $\lambda$ is the ratio of $\Omega$, the perturbation
frequency, to $\omega_0 = ({\cal F G})^{1/2}$, the frequency of
the small-amplitude pendulum oscillations. The parameter $W$ in
the case of $a = b$ is given by

\vspace{-3mm}

\begin{equation}
W = \varepsilon \lambda \left( A_2(\lambda) + A_2(-\lambda)
\right) = \varepsilon {4 \pi \lambda^2 \over \sinh {\pi \lambda
\over 2}} \label{W1}
\end{equation}

\noindent \cite{S98}. Here $A_2(\lambda) = 4 \pi \lambda
{\exp({{\pi \lambda} / 2}) \over \sinh (\pi \lambda)}$ is the
value of the Melnikov--Arnold integral as defined in \cite{C79}.
We use the notation $\varepsilon \equiv a / {\cal F} = b / {\cal
F}$ for the relative amplitude of perturbation. Formula~(\ref{W1})
differs from that given in \cite{C79,LL92} by the term
$A_2(-\lambda)$, which is small for $\lambda \gg 1$. However, its
contribution is significant for $\lambda$ small~\cite{S98}, i.e.,
in the case of adiabatic chaos. Analytical expressions for $W$ at
$a \neq b$ are given in \cite{S00}.

One iteration of map~(\ref{sm}) corresponds to one period of the
pendulum rotation or a half-period of its libration. The motion of
system~(\ref{h}) is mapped by Eqs.~(\ref{sm})
asynchronously~\cite{S98}: the relative energy variable $w$ is
taken at $\varphi = \pm \pi$, while the perturbation phase $\tau$
is taken at $\varphi = 0$. The desynchronization can be removed by
a special procedure~\cite{S98,S00}. The synchronized separatrix
map gives correct representation of the sections of the phase
space of the near-separatrix motion both at high and low
perturbation frequencies; this was found in \cite{S00} by direct
comparison of phase portraits of the separatrix map to the
corresponding sections obtained by numerical integration of the
original systems. This testifies good performance of both the
separatrix map theory and the Melnikov theory, describing the
splitting of the separatrices.

\section{Analytical estimating the width maxima}

Let us present, for clarity, the resulting formulas of the method
\cite{S98} for estimating the locations and heights of the peaks
(Eqs.~(8), (9), and (10) in \cite{S98}).

According to \cite{S98}, the approximate condition  for the
tangency of the unperturbed separatrix of a marginal integer
resonance to the border of the primary chaotic layer
(equivalently, the approximate condition for the maximal energy
jumps in the near-separatrix motion) is given by the equation

\begin{equation}
W(\varepsilon, \lambda) = W_\mathrm{t}^{(m)}(\lambda) ,
\label{tc}
\end{equation}

\noindent where $W(\varepsilon, \lambda)$ and
$W_\mathrm{t}^{(m)}(\lambda)$ are expressed as follows. The
expression for $W(\varepsilon, \lambda)$ is determined by the
choice of the system. For the pendulum with the harmonically
driven point of suspension it is given by Eq.~(\ref{W1}):

\begin{equation}
W(\varepsilon, \lambda) = \varepsilon {4 \pi \lambda^2 \over
\sinh {\pi \lambda \over 2}}.
\label{W}
\end{equation}

\noindent The expression for $W_\mathrm{t}^{(m)}(\lambda)$ does
not depend on the choice of the system. In the pendulum model of
the marginal resonance it has the universal form:

\begin{equation}
W_\mathrm{t}^{(m)}(\lambda) = {32 \over \lambda^3}
    \left[ ( 1 + \lambda^2 )^{1/2} - 1 \right]^2
    \exp \left( - {2 \pi m \over \lambda} \right) ,
\label{Wt}
\end{equation}

\noindent where $m$ is the order of the marginal resonance (see
details in \cite{S98}).

The location $\lambda = \lambda_m$ of an $m$th peak at any value
of $\varepsilon$ can be found by solving the functional
equation~(\ref{tc}) with respect to $\lambda$ at any $\varepsilon$
and $m \geq 1$ numerically. This can be easily done, e.g., in the
Maple computer algebra system \cite{C93}. The extreme value of the
relative energy during the energy jump is given by

\begin{equation}
w_m = \pm \left[ 64 \exp \left( - {2 \pi m \over \lambda_m} \right)
              - \lambda_m W_\mathrm{t}^{(m)}(\lambda_m) \right] .
\label{wextr}
\end{equation}

Recently Soskin and Mannella \cite{SM09} presented a theoretical
method for calculation of maximal width of the separatrix chaotic
layer, which is suitable for a wide class of the periodically
perturbed one-degree of freedom Hamiltonian systems. Their theory
describes the shape of the resulting peak in the ``frequency of
perturbation~--- relative energy'' coordinates, its location in
the frequency of perturbation and its height in the relative
energy. Besides, it provides a general approach in the framework
of Soskin and Mannella's classification of the periodically
perturbed one-degree of freedom Hamiltonian systems. The pendulum
model was not used in \cite{SM09} for description of the marginal
resonance. Instead, a transition from a discrete mapping to
continuous regular-like representation of the motion in some limit
and an analysis of the marginal resonance Hamiltonian were used to
describe the perturbed border of the layer. Note that the system
parameters are designated differently in \cite{SM09} as compared
to our designations in \cite{S98}: (i) the relative frequency of
perturbation is $\lambda$ in \cite{S98}, but $\omega_f$ in
\cite{SM09}, so that $\lambda = \omega_f$; (ii) the relative
amplitude of perturbation is $\varepsilon$ in \cite{S98}, but
$h/2$ in \cite{SM09}, so that $\varepsilon = h/2$.

The formula for the location of the $m$th peak $\lambda_m$ derived
in \cite{SM09} in the adiabatic approximation $h \ll 1$ is

\begin{equation}
\lambda_m \approx - \frac{2 \pi m}{\ln \frac{h}{4}} =
 - \frac{2 \pi m}{\ln \frac{\varepsilon}{2}}
\label{lamS09}
\end{equation}

\noindent (Eq.~63 in \cite{SM09}), and the corresponding
theoretical value of the maximal half-width of the chaotic layer
is

\begin{equation}
| w_m | \approx 2 (4 e + 1) h = 4 (4 e + 1) \varepsilon \approx
23.75 h \label{wehS09}
\end{equation}

\noindent (Eq.~72 in \cite{SM09}).
The limit $h \to 0$ in the description of the peaks is adiabatic,
because $\lambda_m \to 0$ if $h \to 0$.

\section{Computing the width maxima}

At each value of $\lambda$ the half-width can be measured by two
methods, as described in \cite{S08}. The first one was proposed in
\cite{C78,C79} and developed and extensively used in \cite{V04}.
It is based on calculation of the minimum period $T_\mathrm{min}$
of the motion in the chaotic layer. The half-width is determined
by the formula~\cite{C78,C79,V04}:

\vspace{-3mm}

\begin{equation}
w_\mathrm{b} =  32 \exp(- \omega_0 T_\mathrm{min}). \label{wb1}
\end{equation}

\noindent The minimum period corresponds to the maximum energy
deviation from the unperturbed separatrix value. This formula
directly follows from the second line of Eqs.~(\ref{sm}).

The second method consists in the direct continuous measuring of
the relative energy deviation from the unperturbed separatrix $w =
{H_0 \over {\cal F}} - 1$ in the course of integration, and fixing
the extremum one.

In this work, I compute the width of the chaotic layer near the
separatrices of Hamiltonian~(\ref{h}) directly, i.e., by the
second method. The integration of the equations of motion, given
by Hamiltonian~(\ref{h}), has been performed by the integrator by
Hairer et al.~\cite{HNV87}. It is an explicit 8th order
Runge--Kutta method due to Dormand and Prince, with the step size
control.

The integration time interval is chosen to be $10^4$, in the units
of periods of perturbation. Each unit is divided in $10^5$ equal
segments; the trajectories have been output at the end of each
segment, to provide, with such time resolution, the calculation of
the time period and the relative energy deviation of the motion.
Further increasing the integration time interval or decreasing the
length of the segments have been checked to leave the estimates of
the width unchanged within 3--4 significant digits; i.e., the
estimates are saturated enough.

In Table~1, I present the numerical-experimental estimates of
$\lambda_1$ and $| w_1 | / h$, obtained in this way, i.e., by the
direct integrations of the continuous system~(\ref{h}). (It should
be stressed that iterating the separatrix map is not used here.)
The numerical-experimental estimates are given alongside with the
theoretical estimates obtained by the methods \cite{SM09} and
\cite{S98}. The necessary formulas are given above. As follows
from the data in Table~1, a direct comparison of the
numerical-experimental and theoretical results for a set of 5
values of $h$ (namely, $h = 10^{-6}$, $10^{-5}$, $10^{-4}$,
$10^{-3}$, and $10^{-2}$), spanning the whole range of the
perturbation strengths considered in \cite{SM09}, shows that the
theory \cite{S98} performs worse than \cite{SM09} when it is out
of its range of applicability (i.e., when $\lambda_1 \lesssim
1/2$), but it is definitely more accurate than \cite{SM09} when it
is in its range of applicability (when $\lambda_1 \gtrsim 1/2$).
Judging by the absolute deviation of the theoretical values of
$\lambda_1$ from its numerically measured values, the theory
\cite{S98} is 4.9 times better than the theory \cite{SM09} in the
case of $h=10^{-2}$, 1.5 times better in the case of $h=10^{-3}$,
and 1.1 times better in the case of $h=10^{-4}$. (The
corresponding deviations are less in the given proportions.)
Judging by the absolute deviation of the theoretical values of $|
w_1 | / h$ from its numerically measured values, the theory
\cite{S98} is 3.2 times better than the theory \cite{SM09} in the
case of $h=10^{-2}$, 2.0 times better in the case of $h=10^{-3}$,
and 1.4 times better in the case of $h=10^{-4}$. Therefore, the
theory \cite{S98} is definitely more accurate than the theory
\cite{SM09} in the realm of non-adiabatic chaos.

\begin{table}[h!]
\label{table1}
\begin{center}
\caption{The locations and heights of the peaks ($m=1$)}
\begin{tabular}{llllll}
\hline\noalign{\smallskip}
$h$                            & $10^{-6}$ & $10^{-5}$ & $10^{-4}$ & $10^{-3}$ & $10^{-2}$ \\
\hline\noalign{\smallskip}
$\lambda_1$ num., this paper   & $0.420$   & $0.499$   & $0.615$   & $0.800$   & $1.156$ \\
$\lambda_1$ theor. \cite{SM09} & $0.413$   & $0.487$   & $0.593$   & $0.758$   & $1.049$ \\
$\lambda_1$ theor. \cite{S98}  & $0.433$   & $0.515$   & $0.635$   & $0.828$   & $1.178$ \\
$|w_1|/h$ num., this paper   & $26.44$  & $27.27$  & $28.06$  & $28.09$  & $26.69$  \\
$|w_1|/h$ theor. \cite{SM09} & $23.75$  & $23.75$   & $23.75$   & $23.75$   & $23.75$  \\
$|w_1|/h$ theor. \cite{S98}  & $31.68^*$  & $31.51$  & $31.14$  & $30.24$  & $27.60$ \\
\noalign{\smallskip}\hline
\multicolumn{6}{l}{$^*$\rule{0pt}{11pt}
\footnotesize
This value is conditional, because $\lambda_1 < 1/2$.}
\end{tabular}
\end{center}
\end{table}

In fact, the comparative results of the performance of the
theories \cite{SM09} and \cite{S98}, presented in Table~1, are
expectable and natural, because the theory \cite{SM09} was
developed for the case of adiabatic chaos, whereas the theory
\cite{S98} was developed for the cases of intermediary and
non-adiabatic chaos. It is evident that the relatively strong
perturbations $h \gtrsim 10^{-3}$ (corresponding to $\lambda_1
\gtrsim 0.8$) are most common in physical and technical
applications, because the applications usually concern
interactions in multiplets of resonances of comparable strengths;
see e.g.\ \cite{S07}. Therefore, in the range of perturbation
amplitudes most common in applications, the theory \cite{S98}
performs much better than \cite{SM09} in predicting the locations
and heights of the peaks.

From Table~1 it is apparent that, in the given range of $h$, the
predictions of the theory \cite{SM09} are smaller than the
numerical results and the predictions of the theory \cite{S98} are
larger than the numerical results, for both $\lambda_1$ and
$|w_1|/h$. First of all, note that the deviations of the
theoretical $\lambda_1$ and $|w_1|/h$ values from the actual ones
are coherent (i.e., they are both negative or both positive),
because the chaotic layer width, when it is close to a maximal
one, increases with increasing $\lambda$, until the peak value is
achieved and the width sharply drops at the moment of heteroclinic
disconnection of the primary and secondary chaotic layers.

The ``generalized separatrix split'', used in the theory
\cite{SM09} as an addend in the sum representing the total maximal
width of the chaotic layer, is smaller than the width of the
primary chaotic layer when the perturbation is non-adiabatic;
therefore the values of $\lambda_1$ predicted by the theory
\cite{SM09} in the non-adiabatic and intermediary cases are
expected to be smaller than the actual ones. What is more, the
``generalized separatrix split'' is not taken into account at all
in deriving the approximate theoretical estimates in \cite{SM09}.
Thus, on increasing $\lambda$, the moment of heteroclinic
disconnection is somewhat underestimated.

On the other hand, the values of $\lambda_1$ predicted by
\cite{S98} are somewhat overestimated in the given range of $h$.
This is due to the following reason. When the value of $\lambda_m$
corresponds to a slow or intermediary perturbation, the separatrix
cell of the marginal resonance is deformed in comparison with the
lenticular form in the pendulum model (this deformation is evident
in Fig.~2 discussed below). The deformation increases with
decreasing $\lambda_m$. In particular, the ``lower half-width''
(the minimal distance between the resonance center and the lower
branch of the separatrix) of the cell becomes smaller and the
``upper half-width'' greater than the single half-width value
given by the pendulum model, while the total width of the cell
remains rather close to that in the pendulum model. Thus the
theory \cite{S98} uses a somewhat overestimated value of the lower
half-width, and on increasing $\lambda$ in the vicinity of the
peak the moment of heteroclinic disconnection is overestimated.

All the peaks with $m \geq 2$ are located at $\lambda_m >
\lambda_1$, see e.g. Eq.~(\ref{lamS09}). Therefore, in the range
of $h$ covered in Table~1, the peaks of order $m$ higher than 1
are all situated more deeply in the domain of non-adiabatic chaos
than those with $m=1$, and thus the theory \cite{S98} is expected
to dominate over \cite{SM09} in accuracy even to a greater degree.
As my numerical experiments show, this is indeed the case; see
Table~2 (constructed in the same way as Table~1, except that the
case of $m=2$ is considered).

\begin{table}[h!]
\label{table2}
\begin{center}
\caption{The locations and heights of the peaks ($m=2$)}
\begin{tabular}{llllll}
\hline\noalign{\smallskip}
$h$                            & $10^{-6}$ & $10^{-5}$ & $10^{-4}$ & $10^{-3}$ & $10^{-2}$ \\
\hline\noalign{\smallskip}
$\lambda_2$ num., this paper   & $0.853$   & $1.015$   & $1.245$   & $1.603$   & $2.181$ \\
$\lambda_2$ theor. \cite{SM09} & $0.827$   & $0.974$   & $1.186$   & $1.515$   & $2.097$ \\
$\lambda_2$ theor. \cite{S98}  & $0.867$   & $1.029$   & $1.262$   & $1.618$   & $2.188$ \\
$|w_2|/h$ num., this paper   & $28.11$  & $28.01$  & $25.97$  & $23.00$  & $16.27$  \\
$|w_2|/h$ theor. \cite{SM09} & $23.75$  & $23.75$  & $23.75$  & $23.75$  & $23.75$  \\
$|w_2|/h$ theor. \cite{S98}  & $30.01$  & $28.87$  & $26.79$  & $22.92$  & $16.29$ \\
\noalign{\smallskip}\hline
\end{tabular}
\end{center}
\end{table}

For example, let us take $m=2$ and $h = 10^{-2}$. Then the
numerical-experimental values of $\lambda_2$ and $|w_2|/h$ turn
out to be $2.1808$ and $16.27$, respectively. The corresponding
values predicted by the theory \cite{SM09} are $2.0974$ and
$23.75$, and the values predicted by the theory \cite{S98} are
$2.1885$ and $16.29$. Thus, at $h = 10^{-2}$, the theory
\cite{S98} in comparison with \cite{SM09} is $\approx 11$ times
more accurate in predicting $\lambda_2$ and is $\approx 370$ times
more accurate in predicting $|w_2|/h$.

\section{Dynamical validity of the marginal resonance model}

Note that no adiabatic limits were obtained or analyzed in
\cite{S98}, because the theory \cite{S98} was constructed for
another (non-adiabatic) domain of application. The approximation
of marginal resonance in the pendulum model works good in the
range of applicability of the theory \cite{S98}, i.e., in the
non-adiabatic domain (contrary to an opinion expressed in
\cite{SM09}). Indeed, if this model were invalid, than the good
accord between the theory \cite{S98} and the
numerical-experimental data, as demonstrated above, would be
merely a coincidence. But this cannot be the case on the following
reason. In the theory \cite{S98}, the maximal half-width of the
layer is the sum of the half-width of the primary (unperturbed)
chaotic layer and the width of the marginal resonance. The
expression for the width of the primary layer is derived
independently from that for the marginal resonance, and therefore,
if the pendulum model for the marginal resonance were invalid, the
resulting error in the sum could not be compensated by chance.
This means that the good performance of the formulas in estimating
the maximal half-width directly testifies the good performance in
estimating the width of the marginal resonance and, consequently,
its dynamical model.

Let us demonstrate the dynamical validity of the pendulum model
for marginal resonance in the non-adiabatic domain graphically. In
Fig.~1, the phase portrait of the separatrix map

\begin{eqnarray}
     y_{i+1} &=& y_i + \sin x_i, \nonumber \\
     x_{i+1} &=& x_i - \lambda \ln \vert y_{i+1} \vert + c
                   \ \ \ (\mbox{mod } 2 \pi) ,
\label{sm1}
\end{eqnarray}

\noindent is presented at $\lambda = 3$ and $c= 5.55 \mbox{ mod }
2 \pi$. Form~(\ref{sm1}) (adopted in \cite{S98}) of the separatrix
map is equivalent to the classical one~(\ref{sm}). The variables
$x_i \equiv \tau_i + \pi$ and $y_i \equiv w_i / W$ (where $W$ is
given by Eq.~(\ref{W})) are the normalized time and energy,
respectively. The parameter $c$ is given by the following formula
\cite{S98}:

\begin{equation}
c = \lambda \ln {32 \over \vert W \vert} .
\label{c}
\end{equation}

\noindent Solely the chaotic component of the phase space at $y
\geq 0$ is shown in Fig.~1. The phase portrait is synchronized
\cite{S98}: the pairs $x_{i-1}$, $(y_i + y_{i-1})/2$ are drawn
instead of $x_i$, $y_i$, so that the portrait corresponds to a
unified surface of section of phase space. The chosen values of
$\lambda$ and $c$ correspond to the brink of heteroclinic
disconnection between the primary chaotic layer (shown in black)
and the secondary chaotic layer (the chaotic layer of the marginal
resonance; shown in grey): a slightest increase in $c$ separates
the layers, and the width momentarily drops to that of the primary
layer.

From Fig.~1, it is graphically evident that no ``regular''
approximation for the motion, like that in the theory \cite{SM09},
can describe precisely the conditions (the critical values of the
parameters) for the maximal width in the non-adiabatic domain,
because both the primary and secondary layers have substantial
widths. These widths must be calculated and taken into account in
any high-precision theory for estimating the conditions for the
critical heteroclinic connection. Thus it is natural to develop
any theory for estimating the layer width separately for adiabatic
and non-adiabatic cases of perturbation.

It is also evident from Fig.~1 that the separatrix cell of the
marginal resonance is qualitatively described by the theoretical
pendulum cell. The borders of the theoretical cell are depicted by
the continuous curves. They are given by the formula

\begin{equation}
y = y^{(n)} \pm 2 \left( \frac{y^{(n)}}{\lambda} \right)^{1/2}
\cos \frac{x}{2} ,
\label{cell}
\end{equation}

\noindent where $y^{(n)}$ is the location of the center of an
integer resonance of order $n$:

\begin{equation}
y^{(n)} = \exp \frac {c - 2 \pi n}{\lambda} ,
\label{y_n}
\end{equation}

\noindent as can be straightforwardly derived from
Eqs.~(\ref{sm1}).

Of course, on decreasing $\lambda$, the marginal resonance
separatrix cell deforms more and more, and in the adiabatic realm
the pendulum model is hardly applicable for its description. This
is evident from the phase portrait in Fig.~2 (where $\lambda =
0.001$ and $c = 0.0076008  \mbox{ mod } 2 \pi$): here the form of
the separatrix cell is far from the well-known lenticular one
characteristic for the pendulum case.

\section{The bending effect}

As already mentioned above, two methods for computation of the
layer width were used in \cite{S08}. The first one, proposed in
\cite{C79}, is based on calculation of the minimal period of the
motion in the chaotic layer. This minimal period can be converted
to the layer half-width by means of Eq.~(\ref{wb1}). The second
method consists in a direct continuous measuring of the relative
energy deviation from the unperturbed separatrix in the course of
numerical integration, and fixing the extreme deviation. In the
case of the first method, the bending effect is averaged out
\cite{S08}. The theoretical value of the maximal half-width of the
chaotic layer is then given by

\begin{equation}
| w_m | \approx 8 e h = 16 e \varepsilon \approx 21.75 h ,
\label{wehab}
\end{equation}

\noindent instead of formula~(\ref{wehS09}). One can see that the
relative difference between (\ref{wehS09}) and (\ref{wehab}) is
rather small: about 10\% $\approx 1/(4e)$. Compare Figs.~3 and 4
in \cite{S08}: the observed height of the first peak in Fig.~3 is
less by $\approx 10$\% ($\approx 1/(4e)$) than that in Fig.~4. The
reason is that the bending of the layer in the first case is
averaged out, while in the second case it is present, because
different methods (those described above) were used for measuring
the width.

The bending effect is particularly important in the strongly
non-adiabatic case, i.e., at $\lambda \gg 1$. The bending
amplitude in the units of the relative energy $w$, according to
\cite{S98,S00}, is equal to the product of the bending factor
$\delta(\lambda)$ (defined in \cite{S98}) and $W(\lambda,
\varepsilon)$ (given by Eq.~(\ref{W})):

\begin{equation}
W \delta = {W \over \pi} \left\{ \mathrm{Re} \left[ \psi \left( i
{\lambda \over 2} \right) - \psi \left( i {\lambda \over 4}
\right) \right] + {1 \over \lambda^2} - \ln 2 \right\} \sinh {\pi
\lambda \over 2}
\ \underset{\lambda \to \infty}{\sim} \
\frac{8 \varepsilon}{\lambda^2},
\label{dlt}
\end{equation}

\noindent where $\psi(z) = \Gamma^\prime (z) / \Gamma(z)$ is the
digamma function, $i$ is the imaginary unit. The primary layer
half-width is given by

\begin{equation}
w_\mathrm{b} \approx \lambda W \underset{\lambda \to \infty}{\sim}
8 \pi \varepsilon \lambda^3 \exp \left( - \frac{\pi \lambda}{2}
\right) . \label{wb}
\end{equation}

\noindent Comparing Eqs.~(\ref{dlt}) and (\ref{wb}), one finds
that the bending amplitude starts to dominate at $\lambda \approx
7$. This domination is exponential with $\lambda$.

In the theory \cite{S98}, the maximal half-width of the layer is
the sum of the half-width of the primary (unperturbed) chaotic
layer and the width of the marginal resonance. If $\lambda \gg 1$,
the half-width of the primary layer, expressed in the units of the
normalized relative energy $y = w/W$ (where $W$ is given by
Eq.~(\ref{W})), is given by $y_\mathrm{b} \approx \lambda$,
whereas the width of the marginal resonance at the layer border is
given by $\Delta y \approx 2 \pi$ \cite{S04}. Therefore, in the
non-adiabatic limit ($\lambda \to \infty$) the relative (with
respect to the primary layer half-width) jump in the energy due to
the marginal resonance is just $\approx 2 \pi / \lambda$. Hence
the role of marginal resonances in determining the total layer
width diminishes with increasing $\lambda$, and it becomes less
important than that of the primary layer at $\lambda \approx 2
\pi$.

On the other hand, as we have seen, almost at the same value of
$\lambda$ (namely, at $\lambda \approx 7$) the bending effect
starts to dominate in determining the total amplitude of the
energy deviations of the near-separatrix motion. Therefore, solely
the bending effect is important in determining this amplitude at
$\lambda \gtrsim 7$.

\section{The adiabatic case}

Returning to the adiabatic case, let us demonstrate that the
geometrical form of the separatrix cell in that case (illustrated
in Fig.~2) can be described analytically quite easily if one uses
the separatrix map representation~(\ref{sm1}).
Formulas~(\ref{lamS09}) and (\ref{wehab}) can be derived as well,
using this representation. This is made as follows. Let us assume
that the increments of $| x |$ and $| y |$ per iteration in
Eqs.~(\ref{sm1}) are small compared to the total magnitudes of
variation of the corresponding quantities. This is in spirit of an
approximation proposed in \cite{V04} for the separatrix map in
classical form for another problem. For description of the motion
near the separatrix of a marginal integer resonance this
assumption is valid, because the amplitude of variation of $y$ is
much greater than 1, while the increment of $| y |$ per iteration
is less than 1. (Note that in a general situation, when there are
no marginal resonances, such an approximation is invalid; see
discussion in \cite{S08}.) Hence map~(\ref{sm1}) is reduced to the
following differential equation:

\begin{equation}
\frac{d x}{d y} = \frac{- \lambda \ln | y | + c - 2 \pi m}{\sin x}
, \label{diifeqc}
\end{equation}

\noindent where $m$ is the order of the marginal resonance. The
$c$ parameter is determined by the parameters of the original
Hamiltonian system, see Eq.~(\ref{c}). Integrating
Eq.~(\ref{diifeqc}), one has for the guiding curve at $y \geq 0$:

\begin{equation}
(\lambda - \lambda \ln y + c - 2 \pi m ) y = -\cos x + 1 ,
\label{eqc}
\end{equation}

\noindent where the integration constant is set equal to 1, so
that the curve is tangent to the axis $y=0$. This corresponds to
the critical situation: at $y=0$ the motion is stochastized, but a
slightest change of the map parameters can disconnect the curve
from the axis $y=0$, and then the motion is no more stochastized.

One can see that the geometrical form of the separatrix cell of
the marginal resonance in Fig.~2 is described by Eq.~(\ref{eqc})
in a highly accurate way: the analytical curve visually coincides
with the borders of the cell in the phase portrait given by the
separatrix map.

For map~(\ref{sm1}), the unstable fixed point of a marginal
integer resonance of order $m$ is situated at $x=\pi$, $y=\exp[(c
- 2 \pi m)/\lambda]$, and the stable fixed point (center) of the
same resonance is situated at $x=0 \mbox{ mod } 2 \pi$, $y=\exp[(c
- 2 \pi m)/\lambda]$, see Eq.~(\ref{y_n}). Substituting the
coordinates of the unstable fixed point in Eq.~(\ref{eqc}) and
solving the resulting equation with respect to $c$, one has for
the critical value of $c$:

\begin{equation}
c_m = 2 \pi m - \lambda \ln {\lambda \over 2} . \label{ccr}
\end{equation}

\noindent Substituting $c=c_m$ in Eq.~(\ref{eqc}), and solving the
resulting equation with respect to $y$, one finds that at $x=0
\mbox{ mod } 2 \pi$ there are two solutions of Eq.~(\ref{eqc}): $y
= 0$ and $y = 2 e / \lambda$. It is easy to check analytically
that they correspond to two extrema of the $y(x)$ function. (At
$x=\pi$ there is only one solution: $y = 2 / \lambda$.) Therefore,
the maximal value of $y$ is given by

\begin{equation}
y_m = \frac{2 e}{\lambda} . \label{yextr}
\end{equation}

\noindent To connect the obtained values of $c_m$ and $y_m$ with
the values of the parameters of the original Hamiltonian, recall
that $y = w/W$ and $c = \lambda \ln ( 32 / \vert W \vert )$,
where, for the considered Hamiltonian model, $W \approx 8
\varepsilon \lambda$ if $\lambda \ll 1$ (this expression follows
from Eq.~(\ref{W1}); note that a good correspondence of this
expression to the actual amplitude of the separatrix map derived
numerically by integration of the original system was found in
\cite{V04}.) Hence $w_m \approx 16 e \varepsilon$ and $\lambda_m
\approx - {2 \pi m}/{\ln (\varepsilon / 2)}$, in accord with
Eqs.~(\ref{wehab}) and (\ref{lamS09}), respectively.

It is interesting that, as follows from formula~(\ref{yextr}) for
$y_m$ and the fact the $y$ coordinate of the unstable fixed point
is $2 / \lambda$, the relative amplitude of the motion at the
outermost border of the chaotic layer (i.e., the ratio of the
maximal and minimal energies of the motion at the layer border) in
the adiabatic limit is equal to $e \approx 2.718$.

\section{Conclusions}

In this paper, modern theoretical methods for estimating the width
of the chaotic layer in presence of prominent marginal resonances
have been considered in the perturbed pendulum model of nonlinear
resonance. The fields of applicability of these methods have been
explicitly and precisely formulated. The comparative accuracy has
been investigated in massive and long-run numerical experiments.

It has been demonstrated that it is natural to develop any theory
for estimating the layer width separately for adiabatic and
non-adiabatic cases of perturbation. The comparative results of
the numerical performance of the theories \cite{SM09} and
\cite{S98}, given in Tables~1 and 2, unambiguously verify that the
theory \cite{SM09} is suitable in the adiabatic case, whereas the
theory \cite{S98} is suitable in the intermediary and
non-adiabatic cases.

It has been explicitly shown that the pendulum approximation of
marginal resonance works good in the range of applicability of the
theory \cite{S98}, i.e., in the non-adiabatic domain. The role of
marginal resonances in determining the total layer width has been
shown to diminish with increasing the adiabaticity parameter
$\lambda$, and to become less important than that of the primary
layer at $\lambda \approx 2 \pi$. On the other hand, almost at the
same value of $\lambda$ (namely, at $\lambda \approx 7$) the
bending effect starts to dominate in determining the total
amplitude of the energy deviations of the near-separatrix motion.
Therefore, solely the bending effect is important in determining
this amplitude at $\lambda \gtrsim 7$.

In the adiabatic case, it has been demonstrated that the
geometrical form of the separatrix cell can be described
analytically quite easily by means of using a specific
representation of the separatrix map (namely,
representation~(\ref{sm1})). It has been found that the relative
amplitude of the motion at the outermost border of the chaotic
layer (i.e., the ratio of the maximal and minimal energies of the
motion at the layer border) in the adiabatic limit is equal to $e
\approx 2.718$.

The non-adiabatic (and, to some extent, intermediary) case has
been shown to be most actual, in comparison with the adiabatic
one, for the physical or technical applications that concern the
energy jumps in the near-separatrix chaotic motion. The reason is
that even the first peaks (with $m = 1$) appear in the realm of
adiabatic chaos only if the relative strength of perturbation
$\varepsilon$ attains microscopic values: Eq.~(\ref{lamS09})
implies that for $\lambda_1$ to be less than 1/2 the value of
$\varepsilon$ should be less than $2 e^{-4\pi} \approx 10^{-5}$.
In typical applications the strengths of perturbation are much
greater usually. For the second and higher order peaks ($m \geq
2$) to appear in the realm of adiabatic chaos, $\varepsilon$
should be supermicroscopic.

\section*{Acknowledgments}

The author thanks Stanislav Soskin for candid discussions. I am
grateful to anonymous referees for valuable remarks and comments.
This work was partially supported by the Russian Foundation for
Basic Research (project \# 10-02-00383) and by the Programme of
Fundamental Research of the Russian Academy of Sciences
``Fundamental Problems in Nonlinear Dynamics''. The computations
were mostly carried out at the St.\,Petersburg Branch of the Joint
Supercomputer Centre of the Russian Academy of Sciences.

\newpage

\begin{figure}
\centering
\includegraphics[width=0.75\textwidth]{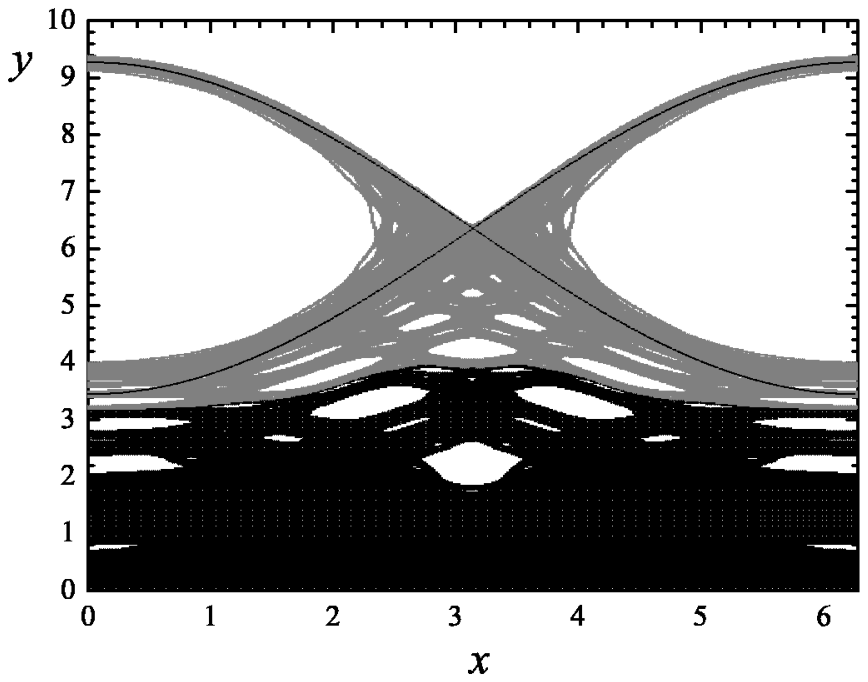}
\caption{The phase portrait of the separatrix map at $\lambda = 3$
and $c= 5.55 \mbox{ mod } 2 \pi$. The theoretical pendulum cell
for the marginal resonance is shown by the continuous curves.}
\label{fig1}
\end{figure}

\begin{figure}
\centering
\includegraphics[width=0.75\textwidth]{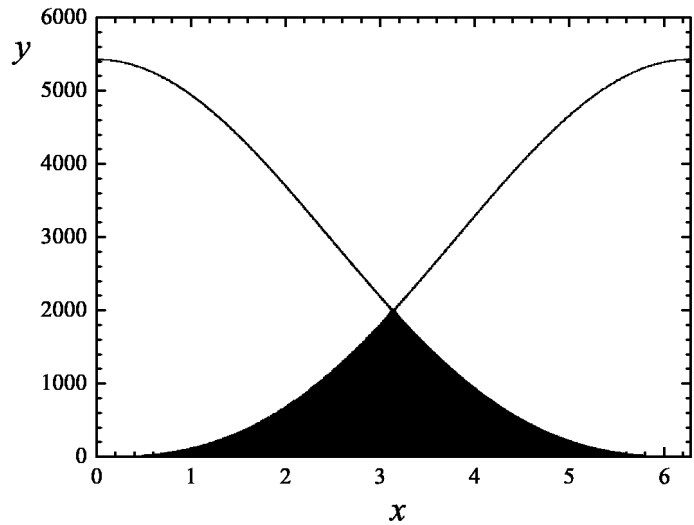}
\caption{The phase portrait of the separatrix map at $\lambda =
0.001$ and $c = 0.0076008 \mbox{ mod } 2 \pi$.}
\label{fig2}
\end{figure}

\end{document}